\documentclass{ws-procs9x6}
\usepackage{graphicx}
\usepackage{epsfig}
\usepackage[figuresright]{rotating}

\newcommand{\cal}[1]{\mathcal{#1}}

\newcommand{\eexp}{{\rm e}}

\newcommand{\vev}[1]{ \langle #1 \rangle }

\begin{document}

\title{The Factorization Method for Simulating  Systems
With a Complex Action} 

\author{J.~AMBJ\O RN}
\address{The Niels Bohr Institute, Blegdamsvej 17, DK-2100 Copenhagen
\O, Denmark} 

\author{K.~N.~ANAGNOSTOPOULOS}
\address{Department of Physics, 
           University of Crete, P.O. Box 2208, GR-71003 Heraklion, Greece}
\author{J.~NISHIMURA}
\address{Theory Division, KEK, Tsukuba,
                 Ibaraki 305--0801, Japan}
\author{J.~J.~M.~VERBAARSCHOT}
\address{Department of Physics
                 and Astronomy, SUNY, Stony Brook, NY 11794, USA}  

%%%%%%%%%%%%%%%%%%%%%%%%%%%%%%%%%%%%%%%%%%%%%%%%%%%%%%%%%%%%%%
% You may repeat \author \address as often as necessary      %
%%%%%%%%%%%%%%%%%%%%%%%%%%%%%%%%%%%%%%%%%%%%%%%%%%%%%%%%%%%%%%

\maketitle

\abstracts{We propose a method for Monte Carlo simulations of systems
with a complex action. The method has the advantages of being in
principle applicable to any such system and provides a solution to the
overlap problem. We apply it in random matrix theory of finite density
QCD where we compare with analytic results. In this model we find
non--commutativity of the limits $\mu\to 0$ and $N\to\infty$ which
could be of relevance in QCD at finite density.}

\section{INTRODUCTION}
There exist many interesting systems in high energy physics whose
action contains an imaginary part, such as QCD at finite baryon
density, Chern-Simons theories, systems with topological terms (like
the $\theta$-term in QCD) and systems with chiral fermions. This
imposes a severe technical problem in the simulations, requiring an
exponentially large amount of data for statistically significant
measurements as the system size is increased or the critical point is
approached. Furthermore, the overlap problem appears when standard
reweighting techniques are applied in such systems and it becomes
exponentially hard with system size to visit the relevant part of the
configuration space. In Ref.\ [\refcite{sign}] it was proposed to take
advantage of a factorization property of the distribution functions of
the observables one is interested to measure. This approach can in
principle be applied to {\it any} system and it eliminates the overlap
problem completely. In some cases it is possible to use finite size
scaling to extrapolate successfully to large system sizes where it
would have been impossible to measure oscillating factors
directly. The method has been applied successfully in matrix models of
non perturbative string theory (IKKT) \cite{sign}, random matrix
theory of finite density QCD (RMT) \cite{rmt} as well as the 2d ${\rm
CP}^3$ model, the 1d antiferromagnetic model with imaginary $B$ and
the 2d compact U(1) with topological charge \cite{Azcoiti}.  In this
paper we present our results for RMT which we study in order to test
the factorization method against known analytical results.  We also
discuss an observed non--commutativity in the limits $\mu\to 0$ and
$N\to\infty$ which maybe relevant to Taylor expansion and imaginary
$\mu$ approaches to the problem of finite density QCD.

The factorization method has the important property that it can be
applied to any system featuring the complex action problem. Let a
system be given by a partition function $Z=\int d{A}\, \eexp^{-S_{
0}}\, \eexp^{i\Gamma}$ and the corresponding phase quenched model
$Z_0=\int d{ A}\, \eexp^{-S_{ 0}}$ where $S=S_0-i\Gamma$ is the action
of the system with its real and imaginary parts. $A$ represents
collectively the degrees of freedom of the model and in our case it
corresponds to a set of $N\times N$ matrices. In case we are
interested in measuring some observable $\cal O$, we consider the
distribution functions $\rho_{\cal O}(x)=\vev{\delta(x-{\cal O})}$ and
$\rho^{(0)}_{\cal O}(x)=\vev{\delta(x-{\cal O})}_0$, where
$\vev{\ldots}_0$ refers to $Z_0$.  Then we define the fiducial system
$Z_{{\cal O},x}=\int d{A}\, \eexp^{-S_{ 0}} \delta(x-{\cal O})$, the
weight factor $w_{\cal O}(x)=\vev{\eexp^{i\Gamma}}_{{\cal O},x}$ and
the distribution $\rho_{\cal O}(x)$ factorizes
%\begin{equation}
%\label{1}
$\rho_{\cal O}(x)=\frac{1}{C}\,\rho_{\cal O}(x)\, w_{\cal O}(x)$ %\, ,
%\end{equation}
where $C=\vev{\eexp^{i\Gamma}}_0$. Then $\vev{{\cal
O}}=\frac{1}{C}\int_{-\infty}^{\infty}dx\,x\,\rho^{(0)}_{\cal
O}(x)\,w_{\cal O}(x)$. The $\delta$--function constraint is
implemented in our simulations by considering the system $Z_{{\cal
O},V}=\int dA \,\eexp^{-S_0}\,\eexp^{V({\cal O})}$ where
$V(z)=\frac{1}{2}\gamma(z-\xi)^2$ and $\gamma, \xi$ are parameters
which control the constraining of the simulation. 
The results are insensitive to the choice of
$\gamma$ as long as it is large enough. Then we have that $w_{\cal
O}(x=\vev{{\cal O}}_{i,V})=\vev{\eexp^{i\Gamma}}_{i,V}$. The
distribution of $\cal O$ in $Z_{i,V}$ has a peak $\bar x$ and the
quantity $V'(\bar x)$ is the value of $f^{(0)}_{\cal
O}(x)=\frac{d}{dx}\ln \rho^{(0)}_{\cal O}(x)$ at $x=\bar x$. The
function $\rho^{(0)}_{\cal O}(x)$ can be obtained by integrating an
analytic function to which  we fit the $f^{(0)}_{\cal O}(x)$ data points.

By applying this method we force the system to sample configurations
which give the essential contributions to $\vev{{\cal O}}$, something
that would be exponentially difficult with system size in the phase
quenched model, eliminating this way the overlap problem. This already
allows us get close to the thermodynamic limit with modest computer
resources. Furthermore we obtain direct knowledge of
$w_{\cal O}(x)$ and $\rho_{\cal O}(x)$ which allows us to understand
the effect of $\Gamma$. This is important for understanding the
properties of the system when $\Gamma$ plays a crucial role. 
%%Using the
%%generic scaling properties of the weight factor $w_i(x)$, one may
%%extrapolate the results obtained by direct Monte Carlo evaluations to
%%larger system size.  Such an extrapolation is expected to be
%%particularly useful in cases where the distribution function turns out
%%to be positive definite.  In those cases we can actually even {\em
%%avoid} using the reweighting formula by reducing the question of
%%obtaining the expectation value to that of finding the minimum of the
%%free energy, which is (minus) the {\it log} of the distribution
%%function.  Here, the error in obtaining the scaling function
%%propagates to the final result without significant magnifications.
%%Therefore, the extrapolation can be a powerful tool to probe the
%%thermodynamic limit from the accessible system size.

\section{RMT OF FINITE DENSITY QCD}

We consider RMT with one quark flavour and zero quark mass
\cite{rmt1}. The model is chosen in order to study the correctness and
effectiveness of the factorization method, since one can compare
results with known analytical solutions even for finite $N$. The
observable we measure is the ``quark number density'' $\nu$ as a
function of the chemical potential $\mu$, and we consider the
distribution functions $\rho_i(x)$, where $i=R,I$ corresponds to the
real and imaginary parts of $\nu$ respectively. Notice that the effect
of $\Gamma$ is dramatic, causing a discontinuous transition in
$\nu$. Our results \cite{rmt} nicely reproduce the exact results
known for finite $N$ and we are able to achieve large enough values of
$N\le 48$ to obtain the thermodynamic limit. In Figure\ 1 we show the
plots of the distribution functions $\rho_{R,I}(x)$ for $\mu=0.2$.
Unfortunately, the function $w_R(x)$ is not positive definite and the
important contributions come from the region where it changes sign. As
expected, we find that finite size scaling does not work as well as in
the case of the IKKT model \cite{sign} (although we obtain agreement
up to order of magnitude for the values of $N\le 96$ that we
explored). We also find it very difficult to explore the critical
region near the phase transition point $\mu_c=0.527\ldots$ for $N>8$
since $|w_i(x)|$ becomes very small.  Since RMT is a schematic model
of finite density QCD, we expect that the factorization method will be
useful to explore the phase diagram of QCD.

\begin{figure}[ht]
%\epsfxsize=10cm   %width of figure - will enlarge/reduce the figures
%\epsfbox{fig3.eps}
%\figurebox{2cm}{3cm}{} %to have a box alone
\centerline{\epsfxsize=3.8in\epsfbox{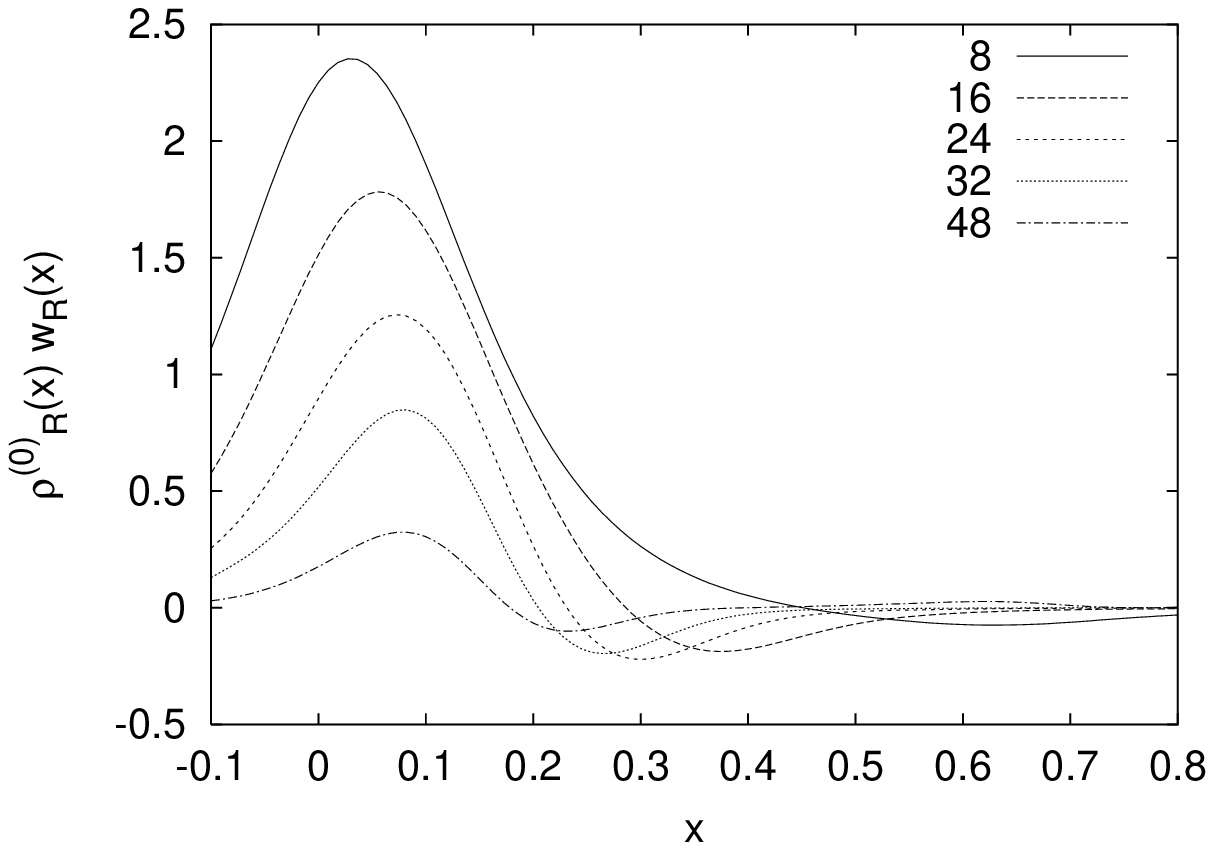}}
\centerline{\epsfxsize=3.8in\epsfbox{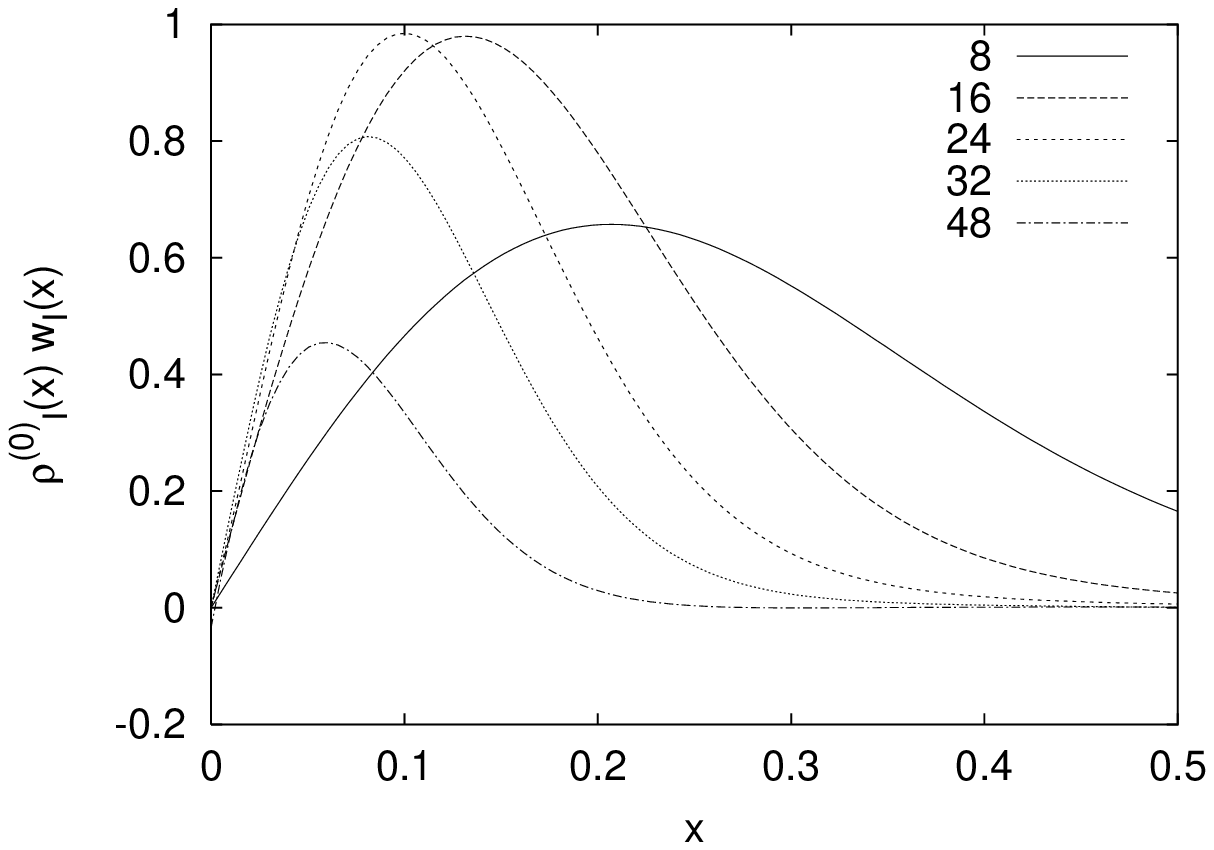}}
\caption{$\rho_R(x)$ and $\rho_I(x)$ for $\mu=0.2$.}
\end{figure}

In our simulations we find that for certain observables the limits
$\mu\to 0$, $N\to\infty$ and $N\to\infty$, $\mu\to 0$ are not
equivalent. In real QCD, the former is easy but not the latter. That
this situation is possible can already be seen at the partition
function level where $Z(\mu,N)=\eexp^\kappa [
1+\frac{(-1)^{N+1}}{N!}\gamma(N+1,\kappa) ]$, $\kappa=-N\mu^2$ is
equal to $1$ and $0$ respectively, but it turns out that the same is
true also for observables like $\frac{\partial}{\partial
\mu}\vev{\nu_R}_0$ and $w_R(x)$. $\vev{\nu_R}$, however, is well
defined in this limit as expected. In the thermodynamic limit
$\vev{\nu_R}_0=\mu$ for $0<\mu<1$. In Figure\ 2 we see that this limit
is approached like $\vev{\nu_R}_0-\mu \sim {\cal O}(1/N)$ but only if
$\mu > \mu_c(N)$. We find that the value of $\mu_c$ is consistent with
$\mu_c^2 \sim 1/N$. A circle with radius $\mu_c = 1/\sqrt{N}$ contains
only one eigenvalue of the matrix on average.  For $N\ll 1/\mu_c^2$ we
find that $\vev{\nu_R}_0=0$. This can be seen clearly from the second
plot of Figure\ 2, where the distributions $\rho^{(0)}_R(x)$ for $N=8$
peak around zero for $\mu<\mu_c$ and their peaks get closer to $\mu$
as $\mu$ becomes larger than $\mu_c$. Therefore
$\lim_{N\to\infty}\lim_{\mu\to 0}\frac{\partial}{\partial
\mu}\vev{\nu_R}_0=0 \ne\lim_{\mu\to
0}\lim_{N\to\infty}\frac{\partial}{\partial\mu}\vev{\nu_R}_0=1$.
Similarly we find that $\lim_{N\to\infty}\lim_{\mu\to 0}
w_R(x)=0\ne\lim_{\mu\to 0}\lim_{N\to\infty} w_R(x)=1$. Details will be
reported elsewhere.
\begin{figure}[ht]
%\epsfxsize=10cm   %width of figure - will enlarge/reduce the figures
%\epsfbox{fig3.eps}
%\figurebox{2cm}{3cm}{} %to have a box alone
\centerline{\epsfxsize=3.8in\epsfbox{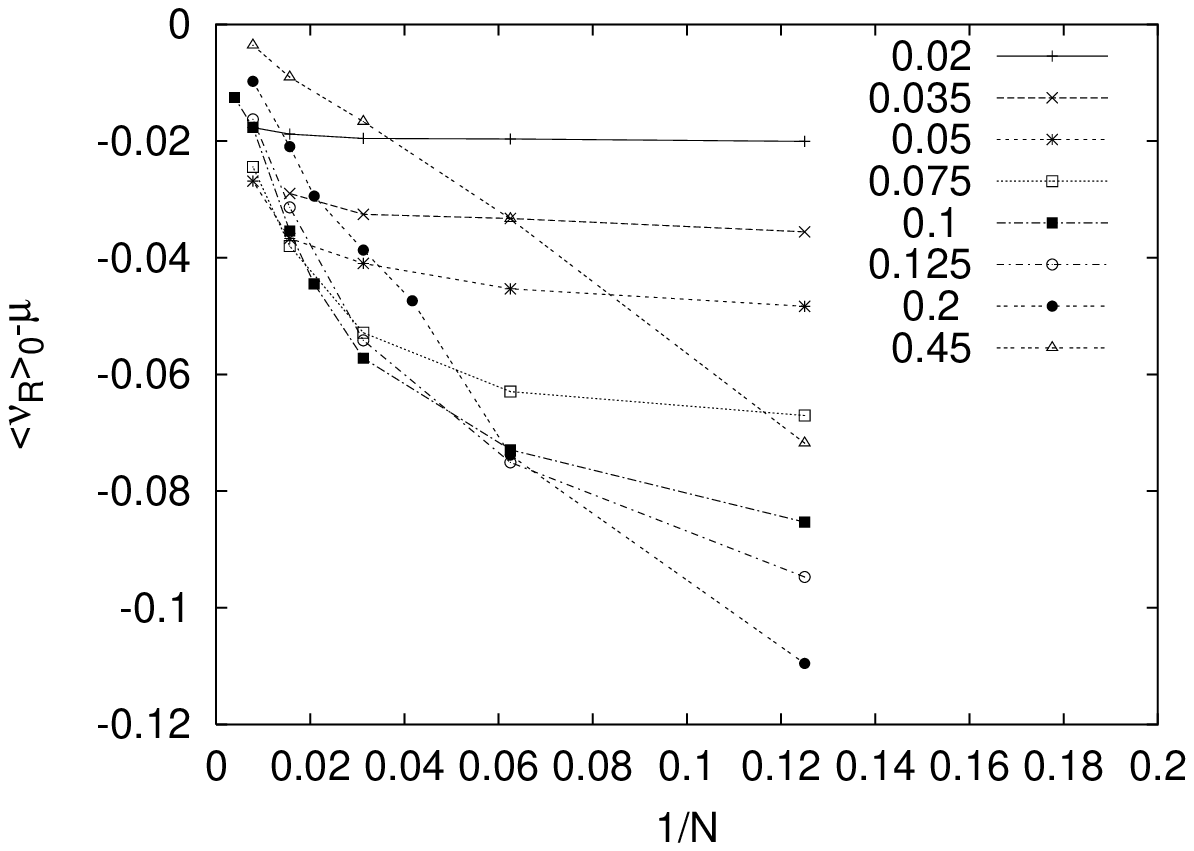}}
\centerline{\epsfxsize=3.8in\epsfbox{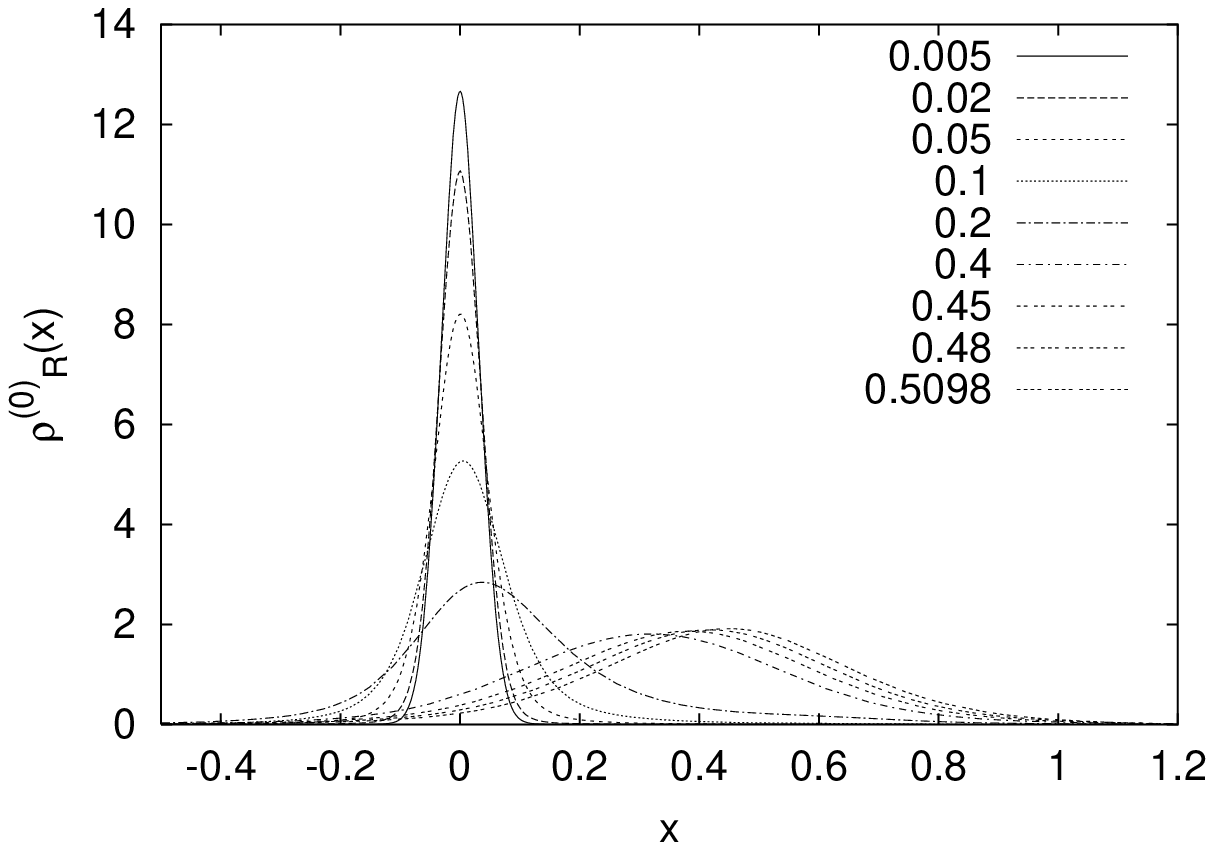}}
\caption{The difference of $\vev{\nu_R}_0$ from its $N=\infty$ value and
  $\rho^{(0)}_R(x)$ for $N=8$ for various values of $\mu$.}
\end{figure}


\begin{thebibliography}{9}
\bibitem{sign}
K.~N.~Anagnostopoulos and J.~Nishimura, Phys.\ Rev.\ D {\bf 66} (2002) 106008.
%%CITATION = HEP-TH 0108041;%%
\bibitem{rmt}
J.~Ambj\o rn et al., JHEP {\bf 0210} (2002) 062.
%%CITATION = HEP-LAT 0208025;%%
\bibitem{Azcoiti}
V.~Azcoiti et al., Phys.\ Rev.\ Lett.\  {\bf 89} (2002) 141601;
hep-lat/0210004. 
%%CITATION = HEP-LAT 0210004;%%
%%CITATION = HEP-LAT 0203017;%%
\bibitem{rmt1}
E.~V.~Shuryak and J.~J.~Verbaarschot,
Nucl.\ Phys.\ A {\bf 560} (1993) 306;
%%CITATION = HEP-TH 9212088;%%
J.~J.~Verbaarschot and T.~Wettig,
Ann.\ Rev.\ Nucl.\ Part.\ Sci.\  {\bf 50} (2000) 343.
%%CITATION = HEP-PH 0003017;%%




\end{thebibliography}
\end{document}